# Sciama's argument on life in a random universe: Distinguishing apples from oranges

Zhi-Wei Wang[1,2] & Samuel L. Braunstein[3]

[1]College of Physics, Jilin University, Changchun, 130012, People's Republic of China;
[2]Department of Physics, University of Warwick, Coventry CV4 7AL, United Kingdom;
[3]Computer Science, University of York, York YO10 5GH, United Kingdom
email: zhiweiwang.phy@gmail.com;
sam.braunstein@york.ac.uk.

Dennis Sciama argued that the existence of life depended on many quantities, the fundamental constants, so in a random universe life should be highly unlikely. However, without full knowledge of these constants, his argument implies a universe that would appear to be 'intelligently designed.'

Dennis Sciama, considered to be one of the fathers of modern cosmology, argued that were our universe random then it would almost certainly have a low probability for life as we know it [1]. In his argument, Sciama assumed that the feature distinguishing different potential universes was the set of specific values taken by the fundamental constants, the underlying physical laws themselves being fixed. We can then envision the human-compatible universes as an 'island' within a 'sea' of more general possibilities. Each point on the island or in the sea describes a unique universe that is described by a distinct set of fundamental constants. The dimensionality of this space of points is naively given by the number of fundamental constants. Thus the human-compatible island of universes corresponds to some shape in a high-dimensional space. The shoreline of the island corresponds to the boundary separating those universes with a chance for human life to form from those where this is impossible. Thus, the shoreline itself will be made up of universes with an exactly vanishing probability for such life. Assuming continuity, as one moves inland, this probability will increase.

This probability landscape is different from the chance of randomly selecting a universe. Because the range of parameters consistent with human life seem to be boxed into a narrow set of values [2], one might expect any smooth measure for randomly selecting universes, to be approximately uniform across the island. However, by the well-known concentration-of-measure phenomenon [3], we can expect that a randomly selected universe will almost certainly be found in the narrow shoreline. In more formal terms, if you paint an $n$-dimensional hypercube with side length $s$, increasing the side length by $\delta$, the total volume of the paint will be $(s+\delta)^n - s^n \geq ns^{n-1}\delta$. But even if $\delta \ll s$, when the number of dimension is sufficiently high that $n \geq s/\delta \gg 1$, the volume of the paint will exceed the volume of the original hypercube.

This is not only true for hypercubes, but for any shape in high dimensions [3]. The volume, and similarly the weight, will be entirely concentrated within a thin layer at the surface. Thus, figuratively, a high-dimensional orange is essentially only its peel. See Fig. 1a-c.

Applied to the high-dimensional island of human-compatible universes, a randomly selected universe will then almost certainly be found in a narrow band on the shore, where by continuity, the probability for life would be expected to be very low. This prediction is in contrast to that of 'intelligent design' where one might expect a universe further inland closer to, or possibly achieving, the greatest chance for human life.

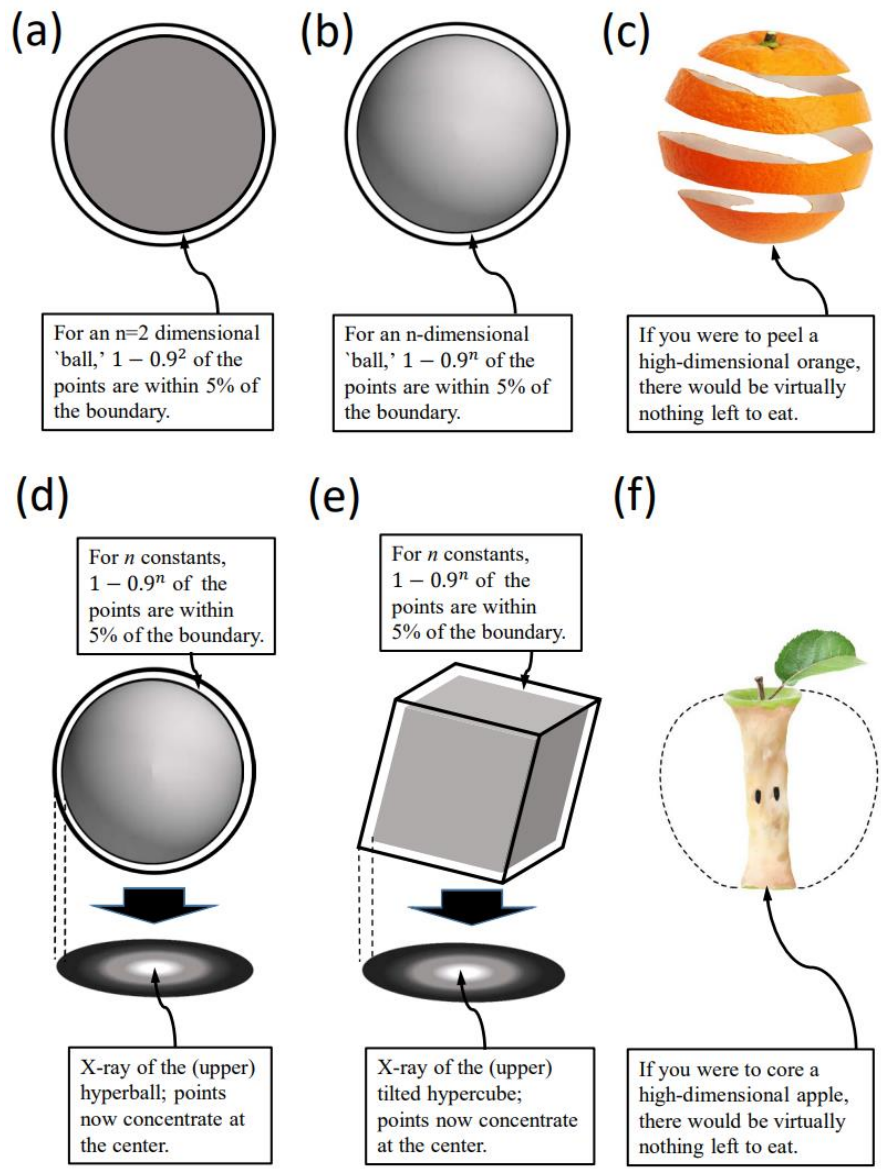

FIG. 1: A counterintuitive consequence of concentration-of-measure phenomena. (a-c): perfect knowledge of the fundamental constants of the universe leads to an "orange" scenario. (a) an $n = 2$ dimensional 'ball' has $1 - 0.9^2 = 0.19$ fraction of the points within 5% of the boundary; (b) an $n = 3$ dimensional ball has a fraction 0.271 within 5% of the boundary; for large $n$, the fraction $1 - 0.9^n$ approaches unity; therefore, (c) if you were to peel a high-dimensional orange, there would be almost nothing left! (d-f): incomplete knowledge of the fundamental constants

leads to an "apple" scenario. (d-e): the projection of a uniformly distributed hyperball or randomly oriented hypercube, shown at the bottom of each panel with high-weight regions in white and the remaining low-weight regions in grey, is well approximated by a narrow gaussian concentrated at the centre; therefore, (f) if you were to core a high-dimensional apple, there would be almost nothing left!

Is this space of universes really high dimensional? In 1936 Eddington counted four fundamental constants [4]. Just a few years ago this count had grown to 26 for the 'standard model' including the cosmological constant for gravity [5]. Today, if we add three neutrino masses, the count would be 29. However, our current model of the universe hardly explains everything, thus it would not be surprising if the total number of the fundamental constants in a complete theory of the universe were much larger.

If our knowledge of the list of fundamental constants were incomplete, we would consider the island and its surrounding sea to be a lower-dimensional space than it actually is – corresponding to an 'X-ray' of the actual $n$-dimensional island onto a lower $m$-dimensional island. Now, if we assume for the sake of simplicity that the island has the shape of a uniform-weight hyperball or randomly oriented hypercube, this X-ray projection would be well approximated with a narrow Gaussian concentrated at the middle of the island (Fig. 1d-e). We argue that this resulting concentration is actually valid for the generic case, for three main reasons [6].

Firstly, the human-compatible island will be formed by those universes whose fundamental constants simultaneously satisfy a series of human-compatible constraints. As mentioned above, the range of parameters consistent with human life is quite small and thus the island itself is in some sense 'small.' Consequently, assuming each constraint is smooth, its action constraining our island should be well approximated locally by a constraining hyperplane in the space of universes. Combined, these hyperplanes yield a human-compatible island with a convex shape.

Secondly, the various correlations and coincidences found among the fundamental constants when determining the human-compatible island's shoreline [2,7,8], suggest that the associated constraining hyperplanes will be tilted with respect to the axes given by the fundamental constants themselves, causing our convex island to have a skewed orientation.

Finally, the projective central limit theorem [9,10] ensures that virtually any projection of such a high $n$-dimensional uniform-weight shape will be well approximated by a Gaussian with variance scaling as $1/n$ with respect to a suitably chosen diameter.

Transferring these considerations to Sciama's argument, one might view his result to be solely that a random universe would lead to a scenario where life as we know it is only barely possible. This 'orange' perspective (Fig. 1 a-c) stands firm and may even explain the apparent scarcity of intelligent life in the universe, potentially resolving Fermi's paradox [1,11]. However, since the island of parameters consistent with any type of lifeform would appear to be significantly larger [12] than that considered solely for the sake of humans, it is possible that life itself may be very common in our universe. This rough-and-ready prediction for a random universe may well be falsifiable within the coming years.

However, presuming that our knowledge of the fundamental constants is incomplete, we have shown that the signature for a random universe can be reversed [6]. Instead, the greatest likelihood would be to find the known constants to be far within the 'projected' human-compatible island of universes, mimicking a universe built by intelligent design to create intelligence [6]. This 'apple' perspective (Fig 1 d-f) would reflect a lack of knowledge.

Currently there is no direct evidence to support the claim that Sciama's statistical signature applies to our universe (outside of consistency with the null results from SETI [13]). However, this observation is in the context of fundamental theories which cannot yet explain everything about our universe, so there is a widely accepted expectation that new physics, along with additional fundamental constants, would be needed. Further, our current best guess for a fundamental theory, string theory, naturally contains a multiverse and hence a random selection mechanism. This reasoned expectation suggests the statistical prediction that many more fundamental constants are yet to be discovered to fully explain our universe.